\documentclass[usenatbib,usegraphicx]{mn2e}

\usepackage{diagbox} % for diagonal line in tables
\usepackage{amsmath}
\usepackage{graphicx}
\usepackage{subcaption}
\usepackage[
pdfauthor={Barbara Ercolano},
pdftitle={PAHs in atmospheres},
pdfstartview=FitH,
linkcolor=blue,
anchorcolor=blue,
citecolor=blue,
filecolor=blue,
menucolor=blue,
urlcolor=blue,
colorlinks=true]{hyperref}
\usepackage{longtable}
\usepackage{aas_macros}
\usepackage{siunitx}
\usepackage{amssymb}
\usepackage{graphicx}  %for images
\graphicspath{{plots/}} 

\usepackage{caption}
\DeclareCaptionLabelFormat{cont}{Figure#1~#2\alph{ContinuedFloat}}
\usepackage{xcolor}

\DeclareSIUnit\Msol{M_\odot}   % Solar mass
\DeclareSIUnit\Zsol{Z_\odot}   % Solar metallicity
\DeclareSIUnit\Gnot{G_0}       % Draine radiation field
\DeclareSIUnit\Myr{Myr}        % Million years
\DeclareSIUnit\kyr{kyr}        % thousand years
\DeclareSIUnit\mag{mag}        % magnitude
\DeclareSIUnit\au{au}          % au

\newcommand\au{\mathrm{au}}

\newcommand{\prodimo}{\textsc{ProDiMo }}

\defcitealias{Picogna2019}{Paper~I}

\title[]{Observations of PAHs in the atmospheres of discs and exoplanets}
\author[Ercolano et al.]{{Barbara Ercolano}$^{1,2}$\thanks{E-mail: ercolano@usm.lmu.de},
       {Christian Rab}$^{1,3}$, 
       {Karan Molaverdikhani}$^{1,2}$, 
       {Billy Edwards}$^{4,5,6,7}$,\and  {Thomas Preibisch}$^{1}$, {Leonardo Testi}$^{8}$, {Inga Kamp}$^{9}$, {Wing-Fai Thi}$^{3}$\\
       $^{1}$ Universit\"ats-Sternwarte, Fakult\"at f\"ur Physik,   Ludwig-Maximilians-Universit\"at M\"unchen, Scheinerstr.~1, 81679 M\"unchen, Germany\\
       $^{2}$ Exzellenzcluster `Origins', Boltzmannstr.~2, 85748 Garching, Germany \\
       $^{3}$ Max-Planck-Institut für extraterrestrische Physik, Giessenbachstrasse 1, 85748 Garching, Germany\\
       $^{4}$ Blue Skies Space Ltd., 69 Wilson Street, London, EC2A 2BB, UK\\
       $^{5}$ AIM, CEA, CNRS, Universit\'e Paris-Saclay, Universit\'e de Paris, F-91191 Gif-sur-Yvette, France\\
       $^{6}$ Department of Physics and Astronomy, University College London, Gower Street, London, WC1E 6BT, UK\\
       $^{7}$ Paris Region Fellow \\
       $^{8}$ European Southern Observatory (ESO), Karl-Schwarzschild-Str 2, D-85748 Garching, Germany \\
       $^{9}$ Kapteyn Astronomical Institute, University of Groningen, Postbus 800, 9700 AV Groningen, The Netherlands}

\newenvironment{itemize*}%
  {\begin{itemize}%
    \setlength{\itemsep}{0pt}%
    \setlength{\parskip}{0pt}}%
  {\end{itemize}}

\begin{document}

\maketitle

\begin{abstract}

Polycyclic aromatic hydrocarbons (PAHs) play a key role in the chemical and hydrodynamical
evolution of the atmospheres of exoplanets and planet-forming discs. If they can survive the planet formation process, PAHs are likely to be involved in pre-biotic chemical reactions eventually leading to more complex molecules such as amino acids and nucleotides, which form the basis for life as we know it. However, the abundance and specific role of PAHs in these environments is largely unknown due to limitations in sensitivity and range of wavelength of current and previous space-borne facilities. Upcoming infrared space spectroscopy missions, such as Twinkle and Ariel, present a unique opportunity to detect PAHs in the atmospheres of exoplanets and planet-forming discs. In this work we present synthetic observations based on conservative numerical modeling of typical planet-forming discs and a transiting hot Saturnian planet around solar type star. Our models show that Twinkle and Ariel might both be able to detect the 3.3 $\mu$m PAH feature within reasonable observing time in discs and transiting planets, assuming that PAHs are present with an abundance of at least one tenth of the interstellar medium value.

\end{abstract}


% ----------------------------------------
% ----------------------------------------
% ----------------------------------------
\section{Introduction} 
\label{sec:intro}

The chemical and hydrodynamical evolution of the gaseous atmospheres of young exoplanets and planet-forming discs is a crucial step for the formation of habitable worlds. Irradiation from the central star plays an important role by kick-starting chemical reactions \citep[see e.g. review from][]{Henning2013} and heating up the atmospheres of discs and planets, which under certain circumstances will become gravitationally unbound and escape the system. Thermally driven winds, also known as photoevaporative outflows, can disperse the planet-forming material in discs and also remove substantial gas from the atmospheres of young planets \citep[see e.g. ][ for recent reviews]{ErcolanoPascucci2017, Owen2019}.  

In this context, large polyatomic molecules known as polycyclic aromatic hydrocarbons (PAHs), play a crucial role in the thermo-chemo-dynamics of gas atmospheres \citep[e.g.][]{Tielens2008}. PAHs are constituted of multiple benzene rings and, due to their high photoelectric yield \citep{BakesTielens1994}, are extremely efficient at converting incoming far-ultraviolet radiation (FUV) into heating for the gas, thus having a strong influence on atmospheric loss rates \citep{Mitani2020,Gorti2009}. PAHs dramatically affect the ionisation level of the gas \citep[e.g.][]{Thi2019} and thus the coupling to magnetic fields, thus also affecting the evolution of the gas in the atmospheres. In the last few years PAHs have become the focus of significant interest in the astrobiology community because of the likely role they play in pre-biotic chemistry, representing an important step towards the formation of amino-acids and nucleotides \citep{Ehrenfreund2007}.  Despite the important role PAHs play in the evolution of planetary and planet-forming disc atmospheres, their abundance in these environments is largely unknown. 

 PAHs distinctive infrared features have been observed everywhere in the interstellar medium (ISM) by space telescopes like the Infrared Space Observatory (ISO) and the Spitzer Space Telescope, since the mid-nineties. The abundance of PAHs (typical size of 50 carbon atoms)  in the ISM is constrained to be $\approx 3\times10^{-7}$ relative to hydrogen nuclei \citep{Tielens2008}. PAHs are also likely to exist in the atmospheres of exoplanets. In our solar system, for example, PAHs have been detected in relatively high concentrations (34 carbon atoms per PAH particle on average) in the  fully evolved nitrogen-rich atmosphere of Saturn’s moon Titan \citep{Dinelli2013,lopez2013large}. However, no search for PAHs in the atmospheres of exoplanets has been possible with current or past instrumentation.

Curiously, PAH detection remains elusive also in the planet-forming discs around young solar-type stars  (T-Tauri stars), while it is commonly observed in discs around more massive stars (Herbig stars) \citep[e.g.][]{Seok2017,Geers2006,Visser2007}. This prompts the question about whether PAHs are destroyed in \mbox{T Tauri} discs, due to the strong X-ray irradiation of their active central stars \citep{Siebenmorgen2010}, while they can survive in the atmospheres of discs around more massive stars that emit less strongly at X-ray wavelengths \citep{Lange2021}. Another possibility, is that PAHs are not detected in \mbox{T Tauri} discs because they have  already been processed into more complex organics through hydrogenation, oxygenation and hydroxylation and have thus lost their infrared signature. The final explanation could be that past and current observational facilities have not had enough sensitivity or have not been able to observe at the appropriate wavelength range in order to detect PAHs in \mbox{T Tauri} discs.

This question remains at the moment unanswered as most disc surveys (e.g. with the Spitzer or Herschel satellites) have not covered a wavelength range including the neutral PAH feature at $3.3\,\mathrm{\mu m}$. They mostly targeted the PAH features at $6.2\,\mathrm{\mu m}$ and $7.6\,\mathrm{\mu m}$, attributed to cations, which are unlikely to dominate in the atmospheres of \mbox{T Tauri} discs. Furthermore, at longer wavelengths the emission is often dominated by silicate emission or absorption, especially for T Tauri stars (see \citealt{Seok2017}). This makes a clear identification PAH features at $6.2\,\mathrm{\mu m}$ and $7.6\,\mathrm{\mu m}$ challenging and should be a lesser issue at $3.3\,\mathrm{\mu m}$. 

Attempts at detecting the $3.3\,\mathrm{\mu m}$ feature from the ground (e.g. with the Very Large Telescope, VLT) have returned limited success (see the high noise level of these observations reported by \citet{Seok2017}, as the Earth atmosphere poses significant limitations at this wavelength. In principle, the $3.3\;\mathrm{\mu m}$ feature could have been detected by the Infrared Space Observatory (ISO) SWS instrument (in orbit for 28 months from November 1995), but the sensitivity was too low to detect it or to set meaningful upper limits for discs around T Tauri stars. 
Fortunately, this situation is likely to change in the near future with the launch of the James Webb Space Telescope\footnote{\url{https://www.cosmos.esa.int/web/jwst}} (JWST) and new facilities like Twinkle\footnote{\url{www.twinkle-spacemission.co.uk}} \citep{edwards_twinkle} and Ariel\footnote{\url{https://arielmission.space}} \citep{tinetti_ariel,tinetti_ariel2}, which will be able to observe in the near-infrared with unprecedented sensitivity. In this work we especially focus on Twinkle and Ariel, because their capabilities for observations of PAHs is unexplored and these missions will have large, dedicated programs for extrasolar science. Furthermore Twinkle and Ariel can provide data complementary to JWST in the mid-infrared, but also in the near-infrared where many planet-forming discs might be too bright for JWST.

The aim of this paper is to present a first investigation of the observability of the 3.3 $\mu$m PAH feature in the atmospheres of nearby planet-forming discs and exoplanets. Through conservative modelling, convolved with instrumental responses of Twinkle and Ariel space telescopes, we obtain synthetic observations of (i) PAH emission spectra in a typical \mbox{T Tauri} disc and (ii) PAH absorption features in transit data of hot-Saturn orbiting G-type star (V=9, K=7.5) for different assumptions of PAH abundance. Our methods and assumptions are described in Section~\ref{sec:methods}. Section~\ref{sec:results} presents our results and discusses them in terms of observability. Finally, Section~\ref{sec:conclusions} contains a short summary of our results. 

\section{Methods} 
\label{sec:methods}

\subsection{PAH cross-sections}
\label{sec:pahcross}
\begin{figure}
\centering
\includegraphics[width=\hsize]{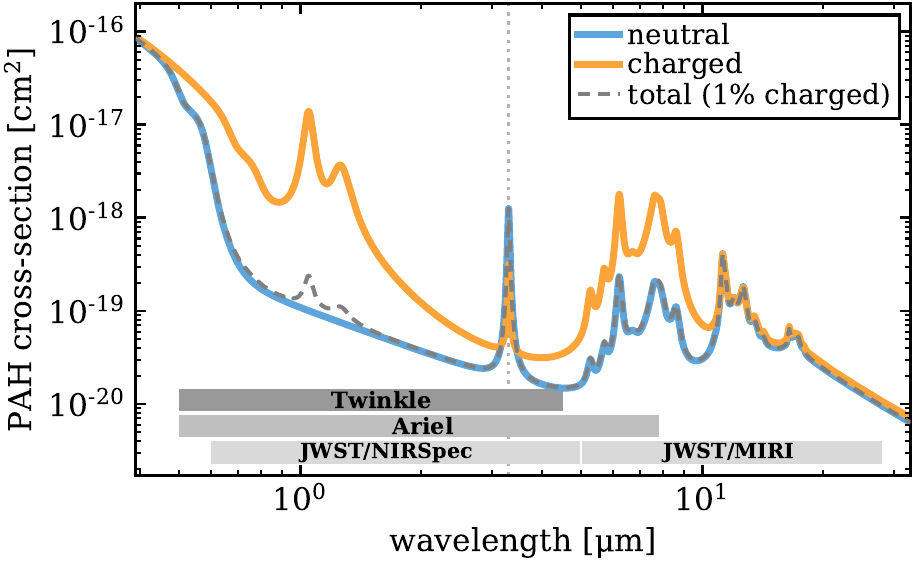}
  \caption{Employed PAH cross-sections for neutral PAHs (blue solid line) and positively charged PAHs (orange solid line). The dashed black line shows the total cross-section assuming a charge fraction of 1\%. The annotated (light) gray bars show the wavelength coverage of the Twinkle, Ariel, JWST/NIRSpec and JWST/MIRI spectrographs. The dotted vertical line marks the $3.3\,\mathrm{\mu m}$ PAH feature.}
     \label{fig:pahcross}
\end{figure}

We calculate the PAH cross-sections following \citet{Li2001} and including updates from \citet{Draine2007} for circumcoronene PAHs consisting of 54 carbon and 16 hydrogen atoms for both neutral and ionized PAHs (for further details also see \citealt{Woitke2016}). The resulting cross-sections are shown in Figure~\ref{fig:pahcross} for the neutral and ionized species and, for a mixture assuming a charge fraction of 1\%. These cross-sections represent optical-properties of "astro-PAHs" and are consistent with astronomical observations in the ISM \citep[see][for details]{Draine2007}. As seen from Fig.~\ref{fig:pahcross} the strength of the $3.3\,\mathrm{\mu m}$ PAH feature is relatively independent of the charge-fraction as the cross-sections for the neutrals and ions are similar, hence we do not expect a strong dependence of the resulting emission on the actual charge fraction of the gas. We use those cross-sections for both the disc and the exoplanet atmosphere models.

\subsection{Calculations of PAH emission spectra from \mbox{T Tauri} discs}
\label{sec:methods_disc}
We model the PAH emission at 3.3 $\mu$m from a disc around a \mbox{T Tauri} star with the radiation thermo-chemical code \prodimo (PROtoplanetary DIsc MOdel, e.g.  \citealt{Woitke2009a,Kamp2010,Thi2011,Rab2018,Rab2020}). \prodimo solves consistently the continuum radiative transfer, the gas and dust thermal balance and the chemical abundances assuming a fixed 2D (radius and height) axisymmetric disc structure for the gas and the dust. The chemical network is described in \citet{Kamp2017} (235 chemical species and 2840 chemical reactions) and includes PAH charge chemistry and adsorption of PAHs onto dust grains.

For the fiducial \mbox{T Tauri} disc model we assume a parameterized flared disc structure with a total disc mass of $M_\mathrm{disc}=0.02\,M_\odot$, a gas-to-dust mass-ratio of 100, an inner radius of $R_\mathrm{in}=0.07\,\mathrm{au}$, a characteristic radius (tapered-edge) of $R_\mathrm{tap}=100\,\mathrm{au}$ (gives an outer disc radius of about $600\,\mathrm{au}$). The scale height of the disc is given by a power-law with a reference scale-height of $10\,\mathrm{au}$ at a radius of $100\,\mathrm{au}$ and a flaring power-law index of 1.15. For further details on this fiducial \mbox{T Tauri} disc model we refer to \citet{Woitke2016}.   

\begin{figure}
    \centering
    \includegraphics[width=\hsize]{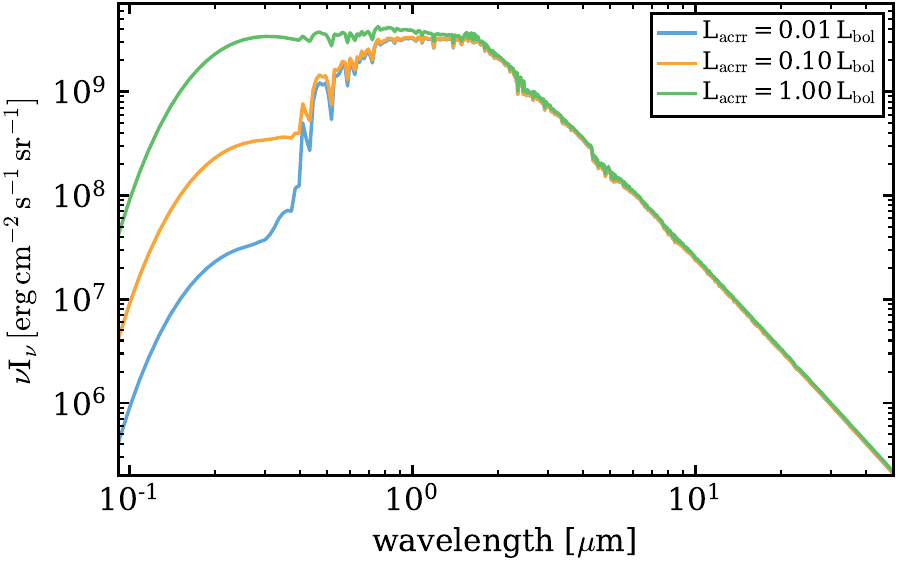}
    \caption{Input stellar spectra for the three discs models with varying accretion luminosity $L_\mathrm{accr}$, representing three mass-accretion rates of $\dot{M}_\mathrm{accr}\approx 10^{-9},\,10^{-8}$ and  $10^{-7}\,M_{\sun} \mathrm{yr}^{-1}$.}
    \label{fig:stellarspec}
\end{figure}
To model the irradiation of the disc by the star in the center we use a PHOENIX stellar spectrum \citep{Brott2005a} for a young solar-like star (age $\approx\!1\!-\!3\,\mathrm{Myr}$) with $M_*=0.7\,M_\odot, L_*=1\,L_\odot$ and $T_\mathrm{eff}=4000\,\mathrm{K}$. To simulate the radiation produced by mass accretion onto the star we use a black body component with $T=12000\,\mathrm{K}$ and consider three different accretion luminosities $L_\mathrm{accr}$ of $0.01, 0.1$ and $1.0\times$ the bolometric luminosity $L_\mathrm{bol}$. The full input spectra are then given by the sum of the accretion component and the photospheric spectrum, and are shown in Figure~\ref{fig:stellarspec}. The chosen accretion luminosities correspond to mass accretion rates\footnote{We assume $L_\mathrm{accr}=\frac{\mathrm{G}M_* {\dot{M}_\mathrm{accr}}}{R_*}$  \citep[e.g.][]{Hartmann1997,Marleau2017}} onto the star of $\dot{M}_\mathrm{accr}\approx 10^{-9}, 10^{-8}$ and  $10^{-7}\,M_{\sun} \mathrm{yr}^{-1}$ and far-UV (ultraviolet) luminosities (definition from  \citealt{Yang2012,Aresu2014}) of $\approx6\times10^{-4}, 6\times10^{-3}$, and $6\times10^{-2}\,L_\mathrm{\odot}$ and are representative of observed mass accretion rates/far-UV luminosities of \mbox{T Tauri} stars \citep[e.g.][]{Yang2012,Alcala2017,Manara2020}. We vary the accretion luminosity because it is especially the far-UV radiation that heats the gas and PAHs.

We use circumcoronene PAHs (see Sect.~\ref{sec:pahcross}) and assume a total PAH abundance in the disc of 0.1 times the typical PAH abundance in the ISM of $3\times10^{-7}$ per hydrogen nuclei \citep{Tielens2008}. The total PAH abundance is fixed in the model and we do not include any formation or destruction processes for the PAH molecules.  
To calculate the PAH emission, the PAHs are included in the wavelength dependent radiative transfer assuming radiative equilibrium to determine the separate PAH and dust temperatures. This is a fast and reasonably accurate method, compared to the quantum heating formalism with stochastic PAH temperature distribution, as shown in \citet{Woitke2016}. 

The resulting SEDs (spectral energy distribution), including the PAH features, are produced via ray tracing assuming a disc inclination of $i=45^\circ$ and a distance to the target of $140\,\mathrm{pc}$. For the PAH cross-sections, used in the radiative-transfer, we assume a fixed charge fraction of 1\% (see Figure~\ref{fig:pahcross}). For a \mbox{T Tauri} disc model like presented here a significantly higher charge fraction is not expected \citep{Thi2019}, and furthermore the $3.3\,\mathrm{\mu m}$ feature does not strongly dependent on the charge fraction (see Sect.~\ref{sec:pahcross}). The modelled synthetic spectra are used as input for the Twinkle/Ariel pipeline for a realistic simulation of the observations (see Sect.~\ref{sec:msynthetic}). 
%We tested one model were we use 10\% charge fraction for the cross-sections and found no significant differences for our results, in particular for the $3.3\,\mathrm{\mu m}$ PAH feature. 

%
\subsection{Calculations of PAH signal in exoplanet transits}
\label{sec:pah_transit}
PAHs are believed to form over a wide range of temperatures \citep[e.g.][]{lopez2013large,zhao2018low}. However, the existance of these molecules is not confirmed on exoplanets. To investigate the feasibility of observing PAHs by the upcoming space facilities, we model the synthetic transmission spectra of a generic exoplanet with the radiative transfer code petitRADTRANS \citep[][]{molliere2019,molliere2020}. As a test case, we assume a typical hot Saturnian planet ($R=0.5\,R_\mathrm{J}$, log(g)=3.0, $T_\mathrm{eq}=1200\,\mathrm{K}$; see e.g. \citealt{Molaverdikhani2019a}) orbiting a solar type star (V=9, K=7.5). In order to isolate the effect of PAH opacity in the atmosphere we assume a hydrogen/helium dominant atmosphere and include the PAHs as the only other opacity source in the atmosphere. For this purpose, no Collision-induced absorption (CIA) or Rayleigh scattering due to H/He was included. For these calculations, we use the same PAH cross-sections as for the disc model (see Sect.~\ref{sec:pahcross} and Figure~\ref{fig:pahcross}). On the Earth, atmospheric PAHs are known to vary in size; ranging from 0.1 to 20 $\mu$m but mostly with a size of a few $\mu$m \citep[e.g.][]{lv2016size}. Such size distribution hints for the importance of Mie scattering contribution at NIR/IR, where the current and future facilities such as JWST, Ariel, and Twinkle will focus on. Consequently, the effect of scattering were also considered using the standard Mie formalism.

%The uncertainty estimations are performed using {\color{orange} ???} code {\color{orange} (Ref?)}, assuming V-mag=9 for the host star, a transit duration of 3 hours, and observing 3 transits in total.
% {\color{red} BILLY could you give more details of the pipeline you used?}

\subsection{Synthetic observations with Twinkle and Ariel}
\label{sec:msynthetic}
To access the detectability of features due to PAHs in both discs and exoplanetary atmospheres, we simulated observations for both Twinkle and Ariel. In each case, we utilised radiometric models which account for a variety of noise sources including shot noise, instrument emission, dark current, read noise and zodiacal light. The instrument simulators are described in depth in \citet{edwards_terminus} and \citet{mugnai_ar} for Twinkle and Ariel, respectively. Additionally, for both observatories, noise due to pointing inaccuracies (i.e. jitter) was imported from dynamical modelling with ExoSim \citep{sarkar_exosim}. Also the relative calibration error across the $3.3\,\mathrm{\mu m}$ PAH feature is included in the instrument simulations.

The spectral resolution in the simulations of Twinkle and Ariel are $\Delta\lambda\approx0.089\,\mathrm{\mu m}$ and $\Delta\lambda\approx0.033\,\mathrm{\mu m}$ at $3.3\,\mathrm{\mu m}$. For comparison the predicted  full-width-half-maximum of the $3.3\,\mathrm{\mu m}$ feature in the model is $\approx0.045\,\mathrm{\mu m}$, but has quite broad wings of $\Delta\lambda\approx0.1\,\mathrm{\mu m}$ at 10\% level of the peak emission (see also Fig.~\ref{fig:pahcross}). To improve on the limited spectral resolution of the instruments we also explore a method that allows to finely sample the region around the PAH feature of interest; the details of this method are described in Appendix~\ref{sec:doublesampling}. 

For each mission, we simulated observations after one hour of integration for the disc model (see Sect.~\ref{sec:methods_disc}), and the uncertainties on the transit depth of a hot-Saturn orbiting the G-type star described in Section \ref{sec:pah_transit} after three observation sequences. We find that both observatories are generally within the photon noise limited regime for our simulations, although, for Twinkle, the dark current is a major noise source in the exoplanet observations considered here.
\section{Results} 
\label{sec:results}
For both the disc case and the planet case we can show that, under the assumptions described in the previous sections, PAHs might be detectable at $3.3\,\mathrm{\mu m}$ by Twinkle and Ariel if they are present at one tenth ISM abundance level ($\approx 3\times10^{-8}$ per hydrogen nuclei) in the atmospheres of typical \mbox{T Tauri} discs or at a mass fraction of $10^{-7}$ (abundance of $\approx3\times10^{-10}$ per hydrogen nuclei) for irradiated planets.

\begin{figure}
    \centering
    \includegraphics[width=\hsize]{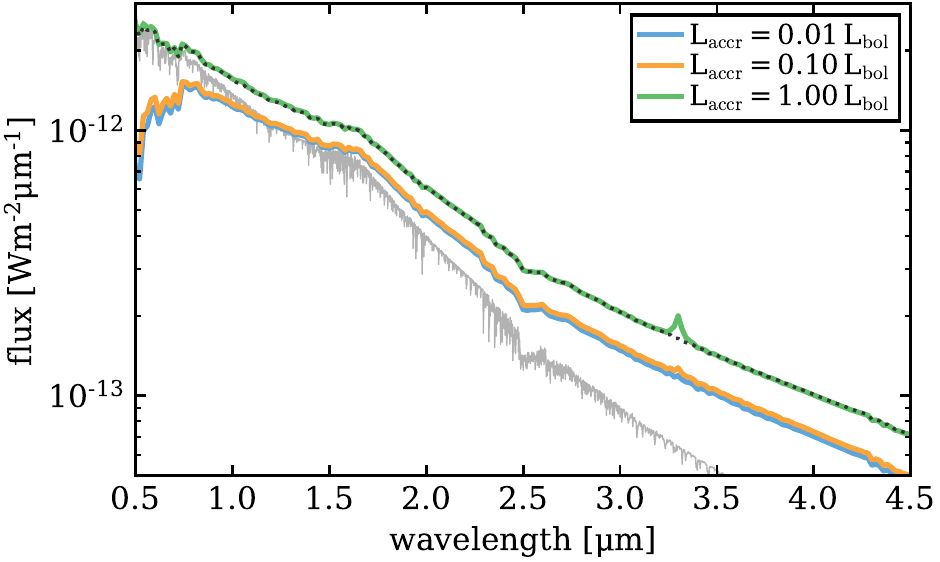}
    \caption{Synthetic disc spectra for the full Twinkle wavelength regime for the three disc models with varying accretion luminosity (solid colored lines). The black dotted line shows the results for a model with $L_\mathrm{accr}=1\,L_\mathrm{bol}$ (where $L_\mathrm{bol}$ is the bolometric luminosity) but without any PAHs in the disc. The grey thin line shows the stellar spectrum for the model with $L_\mathrm{accr}=1\,L_\mathrm{bol}$ as it would be observed if there is no disc.}
    \label{fig:disc_fullspec}
\end{figure}
\subsection{Observability of the $\mathbf{3.3\,\mathbf{\mu m}}$ PAH feature in \mbox{T Tauri} discs}
\label{sec:resultsdiscs}
Figure~\ref{fig:disc_fullspec} shows the SED for the \mbox{T Tauri} disc model from $0.5-4.5\,\mathrm{\mu m}$ (i.e. the Twinkle wavelength regime) as modeled by \prodimo (see Sect.~\ref{sec:methods_disc}), containing PAHs with one tenth of the ISM level abundance. The signal depends strongly on the accretion luminosity in particular the strength of the $3.3\,\mathrm{\mu m}$ PAH feature, which becomes clearly visible for the high accretion luminosity model. Figure~\ref{fig:disc_fullspec} also shows that there are no other easily detectable PAH features, especially at shorter wavelengths the spectral features are mostly dominated by the stellar emission. We therefore focus for the rest of the discussion on the $3.3\,\mathrm{\mu m}$ feature as it is the most promising feature for detection of PAHs in discs with Twinkle/Ariel. For reference we also show the PAH emission at longer wavelengths (e.g. for JWST/MIRI) as predicted by our model in Appendix~\ref{sec:PAHlongwl}.

\begin{figure*}
\centering
\includegraphics[width=0.9\hsize]{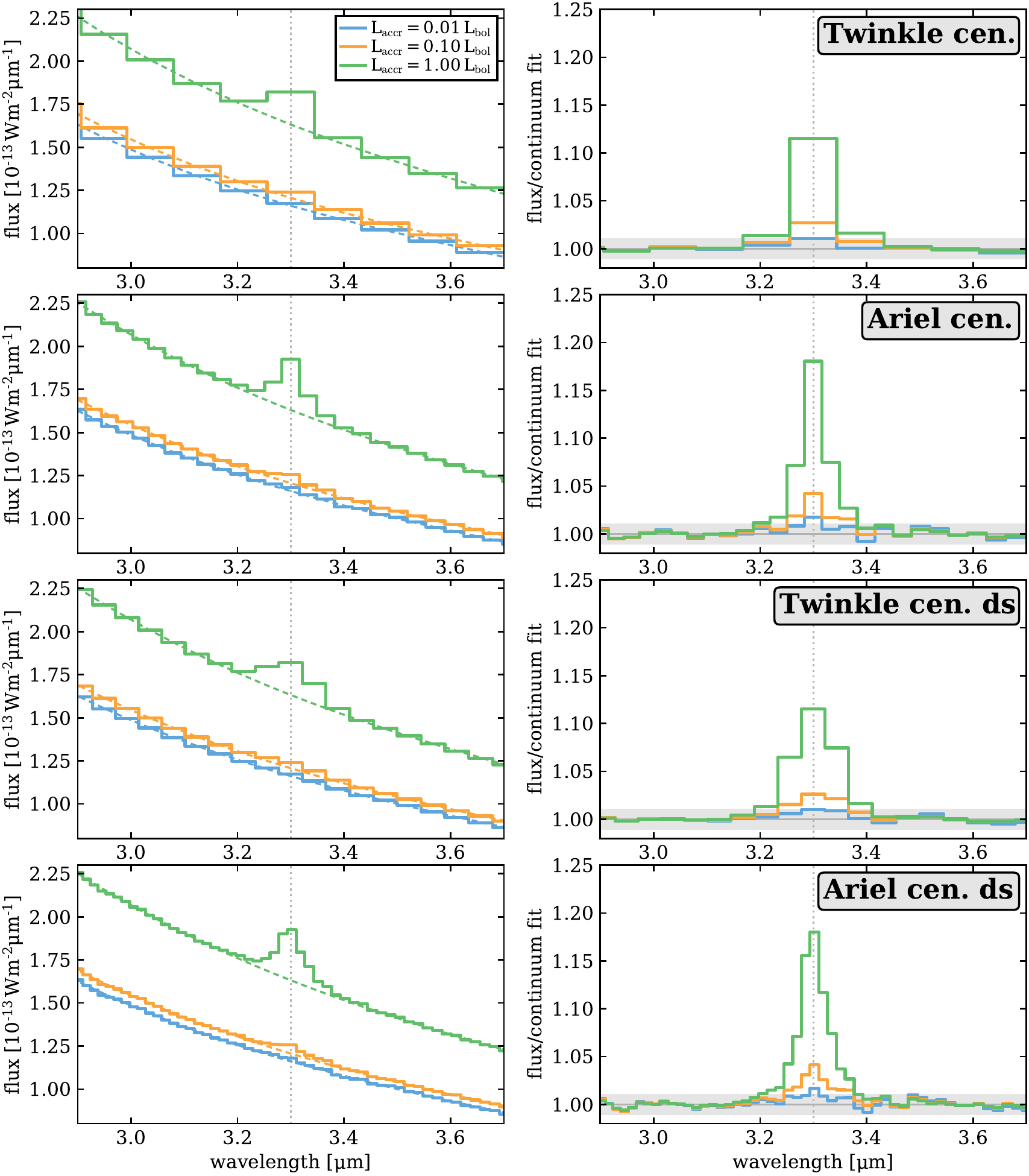}
  \caption{Twinkle and Ariel simulations for the three discs models (colored lines) using spectral bin centering and double sampling (for details see Sect.~\ref{fig:doublesampling}). From \textit{top to bottom} we show the results for Twinkle, Ariel with centered bins (post-fix $\mathtt{cen.}$) at the $3.3\,\mathrm{\mu m}$ PAH feature and with additional double sampling (post-fix $\mathtt{cen.\,ds}$). \textit{Left column}: the simulated spectra for each model (solid colored lines) and the corresponding continuum fit (dashed lines). \textit{Right column}: continuum subtracted spectra (observed flux divided by the continuum fit). The grey area indicates $\pm 1\%$ of the continuum level, for reference. The predicted uncertainties of the instrument simulations for the flux are included, but are not visible as they are smaller than the line thickness. The fluctuations around $\mathrm{flux/continuum \,fit} = 1$ (grey solid line) is a result of the continuum subtraction and not the instrument noise. The vertical dotted line marks the wavelength of $3.3\,\mathrm{\mu m}$ in all panels, for reference.}
  \label{fig:discsim}
\end{figure*}
In Figure~\ref{fig:discsim} we show the results from the Twinkle/Ariel simulations (see Sect.~\ref{sec:msynthetic}) for the $3.3\,\mathrm{\mu m}$ PAH feature using different spectral observing setups. To extract the $3.3\,\mathrm{\mu m}$ PAH feature, we fitted a third order polynomial function around the feature to remove the continuum. The extracted feature is shown in the second column of Figure~\ref{fig:discsim}. The shaded area in this panel indicates $\pm1\%$ of the continuum level. For the models presented here any signal above the 1\% level corresponds to a $10\sigma$ detection of the $3.3\,\mathrm{\mu m}$ PAH feature as long as the signal-to-noise ratio  $\mathrm{SNR}\!\gtrsim\!1000$. This is achievable for both Twinkle and Ariel with observing times less than one hour (see Appendix~\ref{sec:SNR}) for the disc model presented here.    

The first two rows of Figure~\ref{fig:discsim} show the results of the Twinkle/Ariel simulations for the native spectral resolution. We note that we extracted the spectral bins such that, for each instrument, one is positioned at the peak flux of the feature. The last two rows show the expected spectra with a special setup where we use a method we dub ``double sampling", where the datasets are split into two half an hour integrations with different placing of the spectral bins (see Appendix~\ref{sec:doublesampling}). With this setup the spectral data has a lower SNR but the PAH feature is better sampled and therefore the PAH feature peak-flux to continuum ratio is increased. Such a setup is beneficial for extracting the feature, enhancing the chances for a detection and also enabling the observer to ensure any peak in the signal seen is at the expected wavelength. The importance of the spectral resolution and spectral sampling is also apparent from the comparison of the results from Twinkle and Ariel (Figure~\ref{fig:discsim}): Ariel has an approximately two times higher native spectral resolution than Twinkle and thus more easily resolves the feature. While Twinkle time will be dedicated to the study of discs, the bulk of Ariel's will be used specifically for the study of exoplanetary atmospheres. Nonetheless, Ariel will have around 10\% of the mission time set aside for complementary science and thus we hope to motivate the study of discs as part of this program.

The high expected SNRs for both Ariel and Twinkle would in principle allow to extract the $3.3\,\mathrm{\mu m}$ PAH feature for peak-to-continuum flux ratios $< 1.01$ (i.e. at 1\% of the continuum level) with reasonable observing times (i.e. $\approx1\,\mathrm{h}$). This implies that the $3.3\,\mathrm{\mu m}$ feature can be detected with both Ariel and Twinkle with more than $10\sigma$ significance for the models with $L_\mathrm{accr}\gtrsim0.1\,L_\mathrm{bol}$. In the model with $L_\mathrm{accr}=0.01\,L_\mathrm{bol}$ the PAH signal is limited to the central spectral bin (especially for Twinkle), which does not allow for a conclusive identification the feature. It is out of the scope of this paper to fully optimize the extraction of the feature as this work is an exploratory study to assess the capability of Twinkle/Ariel to observe PAHs in \mbox{T Tauri} discs. However, our results indicate that the limiting factor for those telescopes is not the SNR but the spectral resolution (i.e. the feature is smeared out). However, our models allowed to identify a good strategy for placing spectral bins and sampling of the spectrum, which is a promising method to compensate for the expected limited spectral resolution of Twinkle and Ariel.

The NIRSpec\footnote{\url{https://www.cosmos.esa.int/web/jwst-nirspec}} (Near-InfRared Spectrometer, \citealt{Bagnasco2007}) instrument on board of the JWST (James Webb Space Telescope) is also capable of observing the $3.3\,\mathrm{\mu m}$ PAH feature. However, the saturation limit for \mbox{NIRSpec} at $3.3\,\mathrm{\mu m}$ is $\approx \!4.9\times10^{-14}\,\mathrm{Wm^{-2} \mu m^{-1}}$ (full-frame readout, 2-group ramps\footnote{\url{https://www.stsci.edu/jwst/science-planning/proposal-planning-toolbox/sensitivity-and-saturation-limits}}) which is lower than the fluxes of $\approx 1-2 \times 10^{-13}\,\mathrm{Wm^{-2} \mu m^{-1}}$ for the fiducial T~Tauri disc model presented here, indicating that JWST/NIRSpec might not be the ideal instrument for "typical" T-Tauri discs and might be more useful for lower-mass stars or brown dwarfs (a detailed investigation of this aspect is out of the scope of this paper). Twinkle/Ariel will therefore be especially useful for bright targets, in particular targets with high accretion rates and hence strong UV radiation (i.e. high $L_\mathrm{accr}$). Those targets are especially interesting as their strong UV radiation efficiently heats the PAH molecules and therefore strongly enhances the chances of detection of the $3.3\,\mathrm{\mu m}$ PAH feature. Observations with Twinkle/Ariel will also allow to study long/mid-term variability of this interesting near-infrared spectral region due to their long baseline lifetimes (7 and 4 years, respectively) and because they are launched after JWST (expected launch dates are 2021, 2024 and 2029 for JWST, Twinkle and Ariel, respectively).

Disc observations with Twinkle/Ariel will also provide complementary data to mid-IR observations of discs with the Low Resolution Spectrometer (LRS) of the Mid-Infrared Instrument (MIRI\footnote{\url{https://www.jwst.nasa.gov/content/observatory/instruments/miri.html}}, \citealt{kendrew_miri}) of JWST, where the saturation issue is less severe.  Discs observed with MIRI will therefore be ideal targets for follow-up observations with Twinkle and Ariel. Such complementary data will be most relevant for detailed PAH studies (e.g. abundance, composition, sizes) in discs. For example the $3.3\,\mathrm{\mu m}$ PAH feature is likely produced by the C–H stretching modes, whereas the features at mid-infrared wavelengths (see Fig.~\ref{fig:pahcross}) are produce by C–H bending and C–C stretching and bending modes \citep{Draine2007}. Furthermore, detecting PAH emission at various wavelengths opens up the possibility to study PAH abundance as function of disk radius even with spatially unresolved observations, as the short-wavelength emission is expected to emit closer to the star compared to longer wavelengths. 
\begin{figure}
\centering
\includegraphics[width=\hsize]{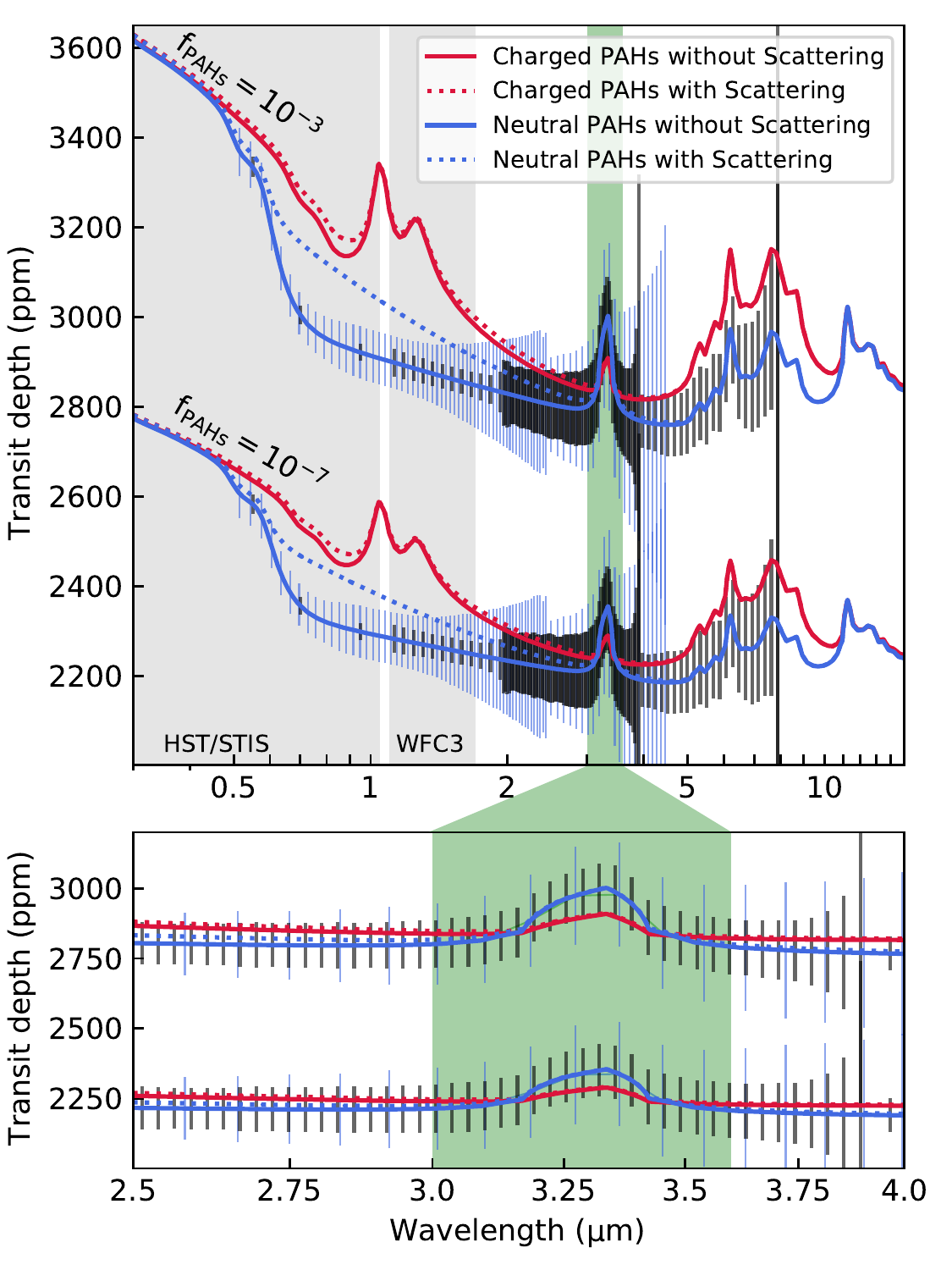}
  \caption{  Top) PAHs signatures in the synthetic transmission spectra of a hot Saturnian (R=0.5$\rm R_{J}$, log(g)=3.0, $\rm T_{eq}$=1200 K) around a G5 type star. The spectra are calculated for charged (solid red) and neutral (solid blue) PAHs and at two mass fractions ($f= 10^{-3}$ and $10^{-7}$, which correspond to abundances of $\approx 3\times10^{-6}$ and $\approx 3\times10^{-10}$ per hydrogen nuclei). The effect of scattering is also shown (dotted lines) for these four cases. As a reference, uncertainties are shown for both Twinkle (blue) and Ariel (black) only in the case of neutral PAHs without scattering (blue error bars), assuming 3 transits with a duration of 3~hr and a stellar V-mag=9. Hubble Space Telescope's STIS and WFC3 wavelength coverage spans over the diagnostic feature of charged PAHs at around 1~$\mathrm{\mu m}$ (gray shaded regions). The diagnostic feature at 3.3~$\mathrm{\mu m}$ is a prominent feature in both charged and neutral PAHs (green shaded region) and hence counts as a promising feature to be investigated by Twinkle. Bottom: zoomed in the top plot around the 3.3~$\mathrm{\mu m}$ feature.}
     \label{fig:planetpah}
\end{figure}

\subsection{Observability of the $\mathbf{3.3\,\mathbf{\mu m}}$ PAH feature in exoplanets transits}

In Figure~\ref{fig:planetpah} we show the expected spectroscopic signatures in terms of transit depth for charged (solid red) and neutral PAHs (solid blue), and at two mass fractions ($f=10^{-3}$ and $10^{-7}$; abundance of \mbox{$\approx3\times10^{-6}$} and \mbox{$\approx3\times10^{-10}$} per hydrogen nuclei). The effect of Mie scattering is also shown with dotted lines for these four cases.

Atmospheres of exoplanets have been successfully characterized by the Hubble Space Telescope (HST). In particular, STIS and WFC3 instruments on board HST have revealed variety of atmospheric constituents on these planets, including water, haze and clouds. Prominent features of the charged PAHs between the 1 to $1.5\,\mathrm{\mu m}$ range fit nicely with the wavelengths coverage of these instruments (shaded gray regions in Figure~\ref{fig:planetpah}). Therefore, they should be already visible in current observation, if they exist in the atmospheres of exoplanets. Our initial inspection of some of the high S/N space-based spectra of exoplanets \citep[e.g.][]{Sing2016,Tsiaras2018} hints against the existence of charged PAHs in the observed exoplanets. Further in-depth retrieval analysis would be needed for setting an upper limit on the charged PAH abundances in exoplanetary atmospheres, which is beyond the scope of current work.

There are a number of features visible in the transit depth of atmospheres containing neutral PAHs. First of all there is a well defined drop-off at approximately $0.6\,\mathrm{\mu m}$ in the absorption cross-section of neutral PAHs. This should be the strongest feature accessible by the current facilities, including HST. Again an initial inspection of some of the best currently available observations reveals otherwise. However, this is expected since scattering effects the shape of spectra and severely washes out this feature as shown by the dotted line in Figure~\ref{fig:planetpah}. The effect of scattering are less prominent at longer wavelengths and hence near and mid-infrared facilities would be geared up with more appropriate tools to investigate PAHs.

The neutral PAH feature at 3.3$\,\mu$m is an example of such near-infrared features. The effect of scattering on this feature is almost negligible and its detection might be feasible through the observation of 3 transits. Due to the magnitude of this feature, even a tracing amount of neutral PAHs in exoplanet atmospheres, e.g. $f=10^{-7}$, might be accessible by the upcoming facilities such as JWST, Twinkle and Ariel. Note that the feature strength changes only by a factor of few between $f=10^{-3}$ and $10^{-7}$ due to the nature of the transit geometry.

The Twinkle mission has dedicated time to constrain PAH abundances in the atmosphere of exoplanets and discs. Uncertainty estimations shown in Figure~\ref{fig:planetpah} suggest a detectable signal should be reachable with 3 transits if the atmosphere contains neutral PAHs, regardless of scattering and for PAHs mass fraction larger than $10^{-7}$. Assuming a transit duration of 3 hours, which is appropriate for a typical hot Jupiter (e.g. HD 209458 b), and allocating 3 hours of out-of-transit baseline, 18 hours observations would be sufficient to achieve the signal shown.

Uncertainties in the transmission spectra are also estimated for the Ariel instruments using the same set of assumptions as described above (these are shown as black errorbars in Figure~\ref{fig:planetpah}). Unsurprisingly, when assuming the same number of observations for each telescope, Ariel provides a better opportunity to explore the noted $3.3\,\mathrm{\mu m}$ feature given the larger collecting area of its primary mirror, which is particularly evident in the noise simulations given the margins currently held on Twinkle's performance. Ariel's extended wavelength coverage to the longer wavelengths would also aid with constraining the features located at wavelengths longer than $5\,\mathrm{\mu m}$. However, Twinkle will provide a better spectral resolution at shorter wavelengths ($<$1.95 $\mu$m) and thus could facilitate the study of possible temporal variations in PAHs signals.

Note that in order to investigate PAH signals, we isolated them in our simulations by excluding all other opacity sources. In reality, additional opacity sources should be expected, e.g. water or methane, which might weaken or mute other molecular features like PAHs \citep[e.g.][]{Molaverdikhani2019a}. Moreover, a variety of atmospheric processes, such as disequilibrium chemistry, might change molecular abundances in the atmosphere of exoplanets and hence spectral signatures \citep[e.g.][]{Molaverdikhani2019b}. Such processes might affect even the hottest exoplanets, where thermochemical equilibrium is expected \citep[e.g.][]{Molaverdikhani2020a}. Clouds might also affect these signatures through their radiative or chemical feedbacks \citep[e.g.][]{2014Natur.505...69K,Molaverdikhani2020b}. Further investigations would be required to quantitatively assess the effect of these processes on the detectability of PAHs in the spectra of exoplanets and under different conditions.
%\begin{figure*}
%\centering
%\includegraphics[width=0.3\hsize]{plots/V11_3_transits.png}
%  \caption{text}
%     \label{fig:twinkleplanet3}
%\end{figure*}

%\begin{figure*}
%\centering
%\includegraphics[width=\hsize]{plots/V11_10_transits.png}
%  \caption{text}
%     \label{fig:twinkleplanet10}
%\end{figure*}

%\begin{figure*}
%\centering
%\includegraphics[width=\hsize]{plots/V11_20_transits.png}
%  \caption{text}
%     \label{fig:twinkleplanet20}
%\end{figure*}
% ----------------------------------------
\section{Conclusions} 
\label{sec:conclusions}

We have performed numerical models of the infrared PAH emission from a typical planet-forming disc and a transiting hot Saturnian planet around a solar-type star. Post-processing our models to obtain synthetic observations with Twinkle and Ariel, we can motivate an observational strategy targeting the $3.3\,\mathrm{\mu m}$ feature, which is typical of neutral PAHs. 

We have shown that assuming a PAH abundance as small as one tenth of the ISM value, both Twinkle and Ariel should be able to detect the $3.3\,\mathrm{\mu m}$  feature in \mbox{T Tauri} discs for typical central star luminosities and accretion properties, within less than one hour observation. 

The 3.3 $\mu$m feature from neutral PAHs should also be detectable in some transiting exoplanets, for abundances one tenth of the ISM value. As an example we have modeled a hot Saturnian planet transiting a G-type star and have shown that the signal to noise ratio required for a firm detection could be made with Twinkle or Ariel within 3 transits which would require fewer than 20 hours of observing time, if we assume a transit duration of 3 hours and 3 hours of out-of-transit baseline.

We note that even non-detections would be extremely valuable as they would allow to constrain stringent upper limits to the PAH abundance in the atmospheric gas, thus providing insights on the relevance of PAH destruction processes in these environments.

\section*{Data Availability}
The data underlying this article are available from the authors upon request.

\section*{Acknowledgements}

We acknowledge the support of the DFG priority program SPP 1992 ``Exploring the Diversity of Extrasolar Planets'' (DFG PR 569/13-1, ER 685/7-1) \& the Deutsche Forschungsgemeinschaft (DFG, German Research Foundation) - 325594231. This reasearch was supported by the Excellence Cluster ORIGINS which is funded by the Deutsche Forschungsgemeinschaft (DFG, German Research Foundation) under Germany's Excellence Strategy - EXC-2094 - 390783311. The simulations have been partly carried out on the computing facilities of the Computational Center for Particle and Astrophysics (C2PAP). CHR is grateful for support from the Max Planck Society. BE is the Project Scientist of the Twinkle space mission at Blue Skies Space Ltd and is a Laureate of the Paris Region fellowship programme which is supported by the Ile-de-France Region and has received funding under the Horizon 2020 innovation framework programme and the Marie Sklodowska-Curie grant agreement no. 945298.

\bibliographystyle{mn2e}
\bibliography{references}

\clearpage

\begin{appendix}

\section{PAH features in the mid-infrared}
\label{sec:PAHlongwl}
\begin{figure}
    \centering
    \includegraphics[width=\hsize]{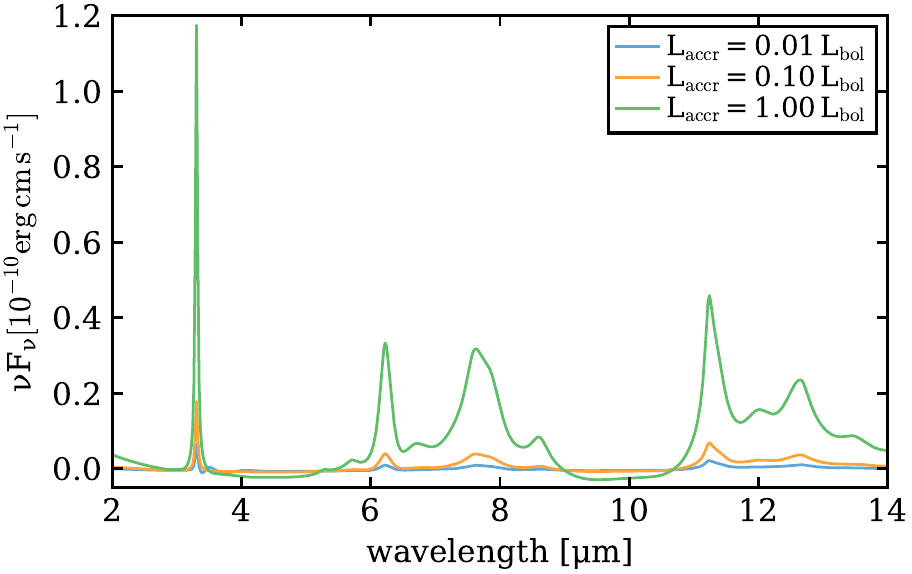}
    \caption{Continuum subtracted PAH spectrum showing also the emission features in the mid-infrared range. This Figure can be directly compared to the results of \citet{Seok2017}.}
    \label{fig:pahmidir}
\end{figure}
For reference we show in Fig.~\ref{fig:pahmidir} the predicted PAH-emission spectrum ranging from $2$ to $14\,\mathrm{\mu m}$ for our three disk models (see Sect.~\ref{sec:methods_disc}). In the mid-infrared the continuum subtraction is more challenging than in at $3.3\,\mathrm{\mu m}$ and furthermore the PAH emission features can be dominated or polluted by silicate emission or absorption (see \citealt{Seok2017}). We therefore chose a simply approach and use the continuum level from models that are identical but do not include any PAHs. We also did not apply any post-processing for simulating a particular instrument as the main purpose is to show that our model predictions in the mid-infrared wavelength regime can be compared to the results presented in \citet{Seok2017}.

In \citet{Seok2017} detailed modeling of PAH emission observations of a large sample of Herbig Ae/BE and a few T~Tauri stars is done. \citet{Seok2017} uses very similar PAH cross-section as are used in this work, but uses more detailed PAH modeling (e.g. varying PAH sizes) and Monte Carlo radiative transfer.  The comparison of their model results for T Tauri stars to our Fig.~\ref{fig:pahmidir} shows that our model predictions are in good agreement with the range of the PAH feature strengths of the the models and the observational data presented in \citet{Seok2017}.

\section{Signal to Noise ratios for the disc models}
\label{sec:SNR}
\begin{figure}
    \centering
    \includegraphics[width=\hsize]{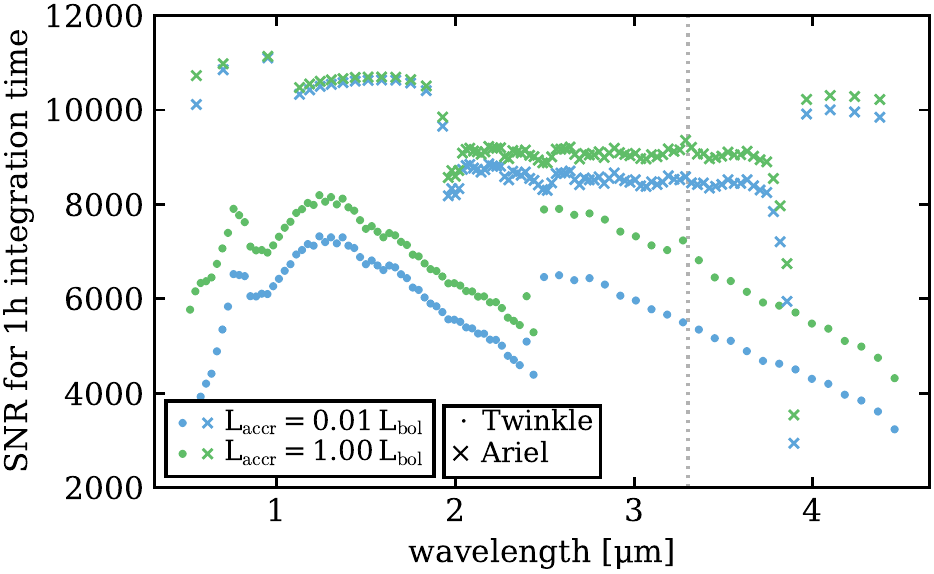}
    \caption{Expected signal to noise ratios (SNR) for Twinkle (dots) and Ariel (crosses) for two different disc models (colors). These are reference values for one hour integration time.}
    \label{fig:SNR}
\end{figure}

In Figure~\ref{fig:SNR} we show the expected signal to noise ratios (SNR) for Twinkle and Ariel for one hour integration time for two of our disc models. The expected SNR are more than sufficient to detect the $3.3\,\mathrm{\mu m}$ PAH feature as simulated in our disc model, and likely shorter observing times are possible (depending on the target). High SNRs are required to extract weak features on top of a strong continuum, which is usually the case in planet-forming discs. Although Twinkle/Ariel were not primarily designed to observe planet-forming discs, the expected high SNRs makes them very interesting options for disc observations in the near-infrared, despite their limited spectral resolution.

\section{Instrument spectral setups }
\label{sec:doublesampling}

\begin{figure}
    \centering
    \includegraphics[width=\hsize]{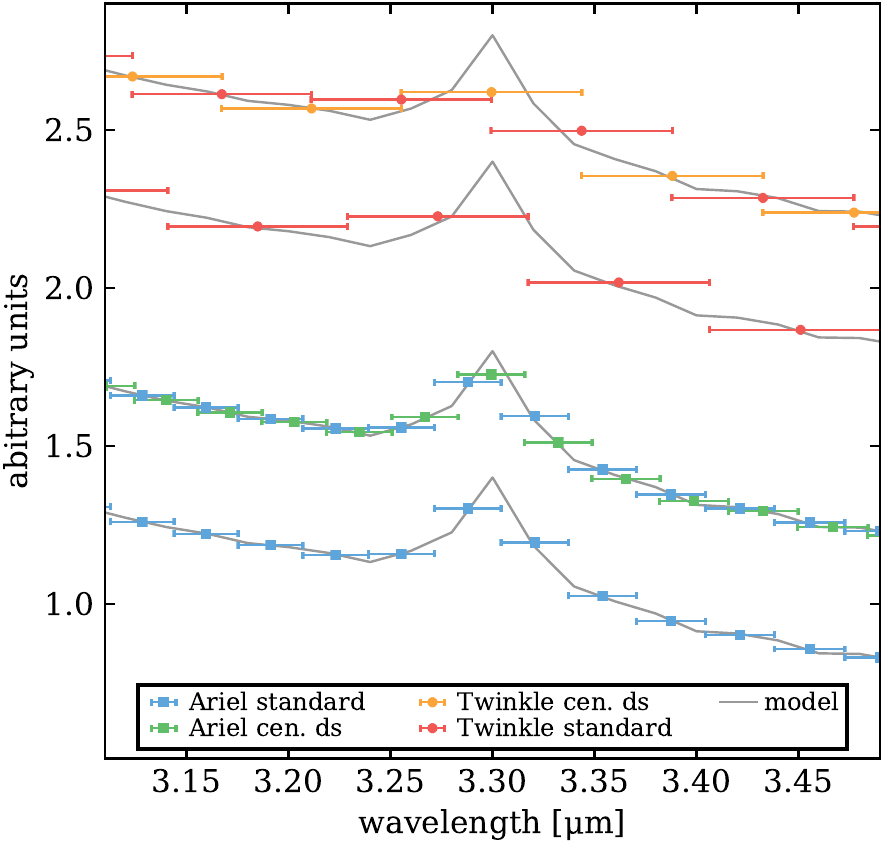}
    \caption{Illustration of the double sampling method. The grey solid lines always shows the same model spectrum (before post-processing with any simulator) and are just shifted for clarity. The errorbars show the results from the instrument simulations. 
    The standard cases (Ariel: blue, Twinkle: red) show the results without any special treatment (i.e. centering). The results using centering (i.e. placing the spectral bin at the peak of the expected feature) and double sampling ($\mathtt{cen.\,ds}$) are shown in two different colors, where each color represents one sampling (i.e. two observing setups with different bin placings are required).}
    \label{fig:doublesampling}
\end{figure}

\begin{figure*}
    \centering
    \includegraphics[width=0.9\hsize]{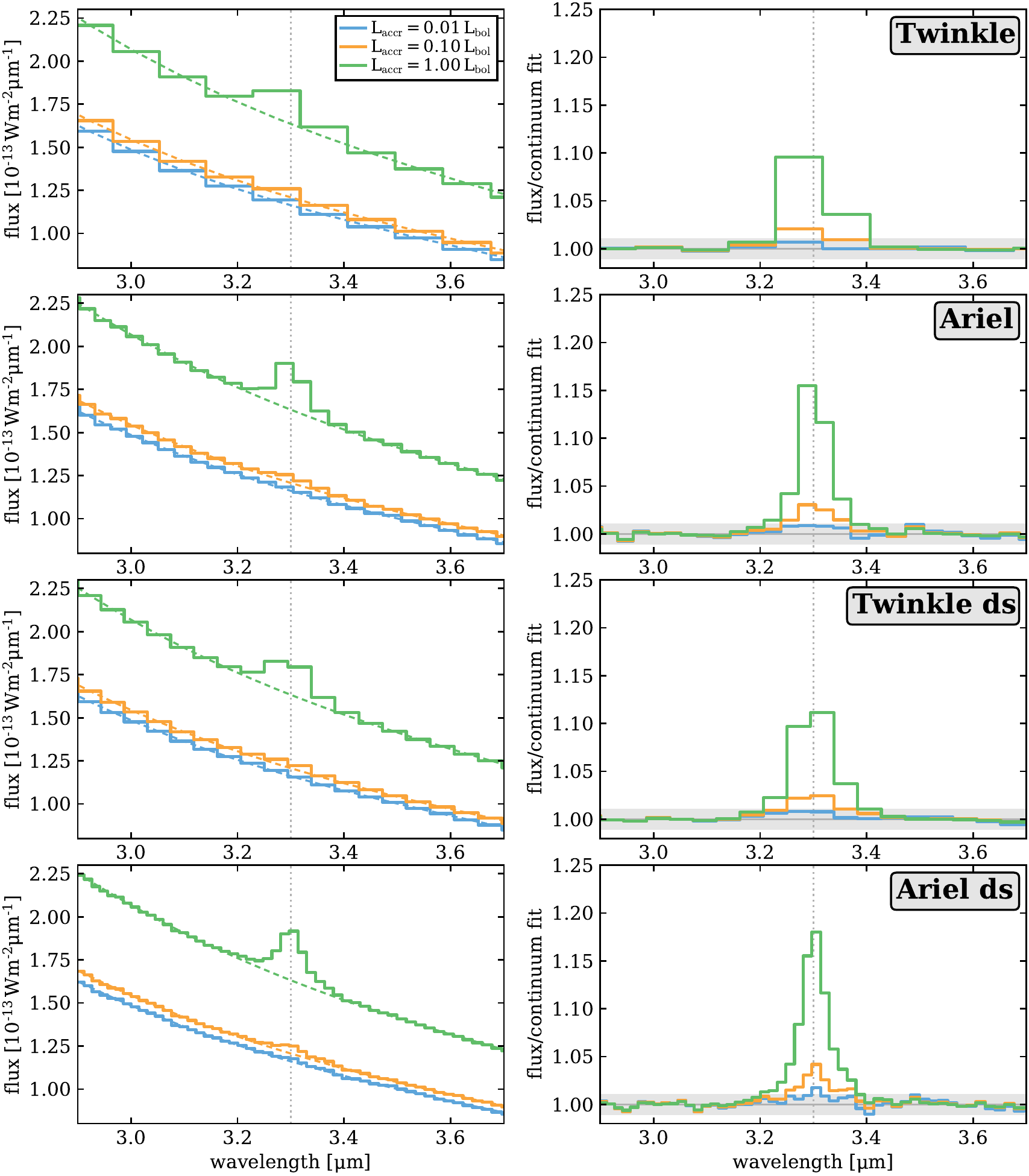}
    \caption{Same as Figure~\ref{fig:discsim} but without using centering of the spectral bins. The top two rows show the results for the native instrument setups of Twinkle and Ariel; the bottom using double sampling without centering (i.e. for cases where the peak of a feature cannot be determined accurately enough).}
    \label{fig:resultsdiscs_native}
\end{figure*}

To improve the peak flux to continuum ratio for the PAH feature, the theoretical PAH spectrum can be used to optimize the placement of the extracted spectral bins of Twinkle and Ariel. In Figure~\ref{fig:doublesampling} we show an example for placing the center of one spectral bin at the wavelength of the expected peak flux of the feature (here $\approx3.3\,\mathrm{\mu m}$) and for double sampling the feature. For double sampling, two observational setups are used with the dataset split into two (i.e. each used for half of the total observing time). For each dataset, the spectral bins were sampled differently, where one sampling should be centered at the peak of the feature and the other offset by half the bandwidth of the bins. The double sampled spectrum is shown by the varying colours in Figure~\ref{fig:doublesampling} and is used in this work for the disc models (Sect.~\ref{sec:resultsdiscs}). For reference, we also show in Figure~\ref{fig:resultsdiscs_native} the results using the native spectral setup of the instruments, and with applying double sampling (model post-fix $\mathtt{ds}$), without centering the bins on the PAH feature at 3.3 $\mu$m.
\end{appendix}
\end{document}